\documentclass[12pt]{article}
\usepackage{geometry}
\usepackage{a4}
\usepackage{graphicx,subfigure}
\usepackage{epsf}
\usepackage{amsmath}
\usepackage{amssymb}
\usepackage{cite}
\newcommand{\be}{\begin{equation}}
\newcommand{\ee}{\end{equation}}
 
 \newcommand{\Rmnum}[1]{\expandafter\@slowromancap\romannumeral #1@}
\newcommand{\bea}{\begin{eqnarray}}
\newcommand{\eea}{\end{eqnarray}}
\begin{document}
\def\A{{\mathbb{A}}}
\def\C{{\mathbb{C}}}
\def\R{{\mathbb{R}}}
\def\s{{\mathbb{S}}}
\def\T{{\mathbb{T}}}
\def\Z{{\mathbb{Z}}}
\def\W{{\mathbb{W}}}
\begin{titlepage}
\title{Thermodynamics and Entanglement Entropy with Weyl Corrections}
\author{}
\date{
Anshuman Dey $^a$\footnote{deyanshu@iitk.ac.in} , Subhash Mahapatra $^b$\footnote{subhmaha@imsc.res.in} , 
Tapobrata Sarkar $^a$\footnote{tapo@iitk.ac.in}
\vskip1.6cm
{\sl $^a$ Department of Physics, \\
Indian Institute of Technology,\\
Kanpur 208016, India.\\
\vskip0.4cm
$^b$ The Institute of Mathematical Sciences, \\
Chennai 600113, India}}
\maketitle
\abstract{
We consider charged black holes in four dimensional AdS space, in the presence of a Weyl correction. We obtain the solution including
the effect of back-reaction, perturbatively up to first order in the Weyl coupling, and study its thermodynamic properties. This is complemented 
by a calculation of the holographic entanglement entropy of the boundary theory. The consistency of results obtained from both
computations is established. 
}
\end{titlepage}

\section{Introduction}
Einstein's general relativity has a deep connection with phenomena of strongly coupled quantum field theory, generally understood via the gauge/gravity
duality, or the AdS/CFT correspondence \cite{Maldacena}, \cite{Klebanov}, \cite{Witten}. According to this duality, the classical gravity solution in
an anti-de Sitter (AdS) spacetime is dual to the large $N$ limit of a strongly coupled gauge theory in one less spacetime dimension, and this duality is 
realized as a form of the holographic principle. For example, to understand a strongly coupled gauge theory at finite temperature, we need a black hole
in AdS space, and the Hawking temperature of the black hole is identified with the temperature of the gauge theory. Since black holes possess temperature
and entropy, it is a natural question to ask whether they show phase transitions like the typical thermodynamical systems in condensed matter physics.
This question was answered long ago by the authors of \cite{HawkingPage} who observed a first order phase transition from the AdS Schwarzschild black
hole to the thermal AdS, known as the Hawking-Page transition in the literatures. 

Since black holes in AdS space exhibit phase transitions just like an
thermodynamical object, the immediate question is, what does this phase transition indicate in the boundary gauge theory. This query was also resolved
in \cite{Witten1}, where it was explained that the Hawking-Page transition is a  holographic dual of the confinement/deconfinement transition in the 
boundary gauge theory. Considering the Reissner-Nordstr\"{o}m AdS (RN-AdS) black hole as a charged solution for the Einstein-Maxwell action with a negative 
cosmological constant, the work of \cite{Chamblin} analyzed the behavior of phase transitions for both the fixed charge and fixed potential ensembles. In fact, in
the fixed charge ensemble, for a range of values of the charge parameter, the horizon radius and hence the black hole entropy exhibits a first order phase 
transition, that resembles the celebrated Van der Waals-Maxwell liquid-gas systems. There exists a critical value of the charge parameter $q_{c}$,
where, the lines of first order phase transition culminate at a second order critical point. 

On the other hand, there is another efficient tool to probe the characteristics of the dual field theory : the entanglement entropy (EE). Simply put, 
if a quantum system is divided into two subsystems $\mathcal{A}$ and $\mathcal{B}$ and we carry out a measurement on $\mathcal{A}$, it would affect 
the measurement on $\mathcal{B}$, provided the two subsystems are entangled. Entanglement entropy is a quantitative measure of this entanglement and it 
explains how strongly the two subsystems are correlated. Since entanglement entropy is related to the number of degrees of freedom of 
the system, there is a lot of interest to probe different phases of a generic quantum field theory via this quantity \cite{Cardy}. In the string theory literature, 
the remarkable conjecture by Ryu and Takayanagi \cite{RyuTakayanagi} offers a method to compute entanglement entropy via holography. 
Using the gauge/gravity duality, the work of \cite{RyuTakayanagi} proposed an elegant formula for the entanglement entropy, which is similar to the Bekenstein-Hawking
formula for the black hole entropy. Because of this similarity, one may wonder whether the entanglement entropy on the boundary field theory can encode 
the dual AdS black hole physics. 

The author of \cite{Johnson} investigated the phase transitions in the dual gravity geometry using the holographic
entanglement entropy (HEE) as a probe. It was shown that the entanglement entropy also reveals a Van der Waals-Maxwell type of phase transition just like
the black hole entropy in a fixed charge ensemble. The value of the transition temperature and the critical charge were seen to match with the results 
obtained by \cite{Chamblin} from direct calculations of black hole thermodynamics. In a recent article, \cite{Caceres} has shown that entanglement
entropy captures the information of the extended thermodynamics of STU black holes in four spacetime dimensions. In a related work \cite{Nguyen}, the 
author examined whether Maxwell's equal area construction holds for the temperature-entanglement entropy plane and the answer was in the affirmative
in a canonical ensemble for the RN-AdS black hole backgrounds. 

It is natural to investigate some of the above phenomena in the presence of extra ``control parameters'' of the theory. Phenomenologically, these are expected to 
enrich the phase diagram and hence might allow for freedom to match with realistic models on the field theory side. Importantly, such terms can arise in a consistent
manner as corrections to a low energy string effective action. For example, one can think of 
studying black hole thermodynamics and its implications on the dual field theory, in the presence of general four derivative interactions over and above 
the Einstein-Maxwell action with a negative cosmological constant. We note here that in \cite{Camanho}, it was shown that such terms can in general
give rise to causality violation, which can be avoided by a mechanism of pair-creation. As we have mentioned, in a top-down approach, these terms are important 
as they naturally arise as quantum corrections to the low energy effective action of string theory. In general, there are a large number of terms that can be
consistently added to a two-derivative Einstein-Maxwell action \cite{Myers}. In four dimensions, it turns out that field redefinitions render a possible eight 
non-zero coupling constants associated to these \cite{Sachdev}. A general analysis involving all these possible eight couplings
is a daunting task, and one does not expect to extract much meaningful physics out of such an analysis. 

We will thus consider turning on a class
of terms that combine the gauge field to the space-time curvature. A particular linear combination of their coupling constants make the action considerably simple,
while providing an extra tuneable parameter that might nonetheless modify the physics non-trivially. To be more precise, as in \cite{Sachdev}, 
we consider here the two derivative Einstein-Maxwell action corrected by a Weyl coupling, which couples the gauge field to the curvature of the bulk spacetime.
This kind of correction term has appeared in \cite{Ritz}, \cite{Sachdev} who  computed the corrections to the conductivity
and the diffusion constant due to a Weyl coupling $\gamma$ and predicted a bound in $\gamma$ from physical
consistency conditions. Recently, the effect of this interaction term on thermalization in a four dimensional boundary theory was studied in detail in \cite{Dey} 
and it was shown that makes the process of thermalization faster, and also points to a number of other non-trivial effects that might be relevant
in a realistic strongly coupled field theory. 

In this paper, we will focus on a Weyl corrected bulk theory in four dimensions. In particular, we consider charged AdS black holes in a fixed charge ensemble. 
The exact solution for this system is difficult to find, and we resort to a perturbative analysis, by treating the dimensionless Weyl coupling $\gamma$ as
a small parameter, and considering the theory up to first order in $\gamma$. 
We first study the thermodynamics of such a theory, and then analyze aspects of its entanglement entropy. The former is straightforward, and we derive 
corrected forms of the thermodynamic variables that are shown to satisfy the first law
of thermodynamics in the presence of a Weyl correction. As far as the entanglement entropy is concerned, we can not use the Ryu-Takayanagi 
formula to compute the entanglement entropy since, the formula holds for a static background with Einstein gravity as the bulk action. For a generic higher 
derivative gravity theory, \cite{Myers1}, \cite{Dong} developed a prescription to compute the entanglement entropy, and we resort to such an analysis. 
In particular, \cite{Dong} proposed a general formula for the entanglement entropy which gets the leading contribution 
from Wald entropy \cite{Wald} and the subleading contributions from the extrinsic curvatures. We will argue that in our case, the entanglement entropy 
would get nonzero contribution only from the Wald entropy. 

In particular, we consider here a global AdS geometry with Weyl correction, having boundary 
$\mathbb{R}\times S^2 $ and on $S^2$ we consider a subsystem of small volume with respect to its complement and calculate the HEE
of that subsystem. This is in order to avoid the regime where the entanglement entropy is dominated by the thermal entropy.
We show explicitly how the HEE of the boundary field theory correctly encodes all the aspects of black hole physics with Weyl corrections. 
The critical behaviour of the entanglement entropy near a second order phase transition point will also been presented in detail, along with the computation of 
the critical exponent. 

The paper is organized as follows : In section 2, we construct the black hole solution with both spherical and planar horizon topology up to linear order
in the Weyl coupling constant. In section 3, we analyze the thermodynamics of our solution, and explain the resemblance with Van der Waals liquid-gas
system, and verify the equal area law. In section 4, we compute the holographic entanglement entropy of the boundary field theory and revisit 
all the results we get from section 3. In section 5, we conclude by summarizing our main results.  

\section{Black hole solution with Weyl correction}
In this section, following, we consider a four-dimensional gravity with a negative cosmological constant, coupled to a $U(1)$ gauge field 
`$A$' by the following two and four-derivative interactions :
\begin{eqnarray}
&&\textit{S} = \frac{1}{16 \pi G_4} \int \mathrm{d^4}x \sqrt{-g} \ \ \bigl[R+\frac{6}{L^{2}}
-\frac{1}{4}\textit{F}_{\mu\nu}\textit{F}^{\mu\nu}+ L^2 (c_1\textit{R}_{\mu \nu \rho \lambda}
\textit{F}^{\mu\nu}\textit{F}^{\rho\lambda} \nonumber\\
&&+c_2 R_{\mu\nu}\textit{F}^{\mu}\hspace{0.1mm} _{\rho}\textit{F}^{\nu\rho} + c_3 R F_{\mu\nu}F^{\mu\nu})\bigr] ,
\label{actiongeneral}
\end{eqnarray}
where $F=dA$ and $F_{\mu\nu}F^{\mu\nu}$ is the familiar two derivative interaction term. The coefficients $c_1$, $c_2$ and $c_3$ define dimensionless
coupling constants for the four derivative interaction terms which couple two derivatives of the gauge field to the spacetime curvature. 
$L$ is the AdS length and it is related to the cosmological constant $\Lambda$ by, $\Lambda=-{3\over L^2}$, $G_4$ is the gravitational constant in four
dimension. 

Following \cite{Ritz}, \cite{Sachdev}, \cite{Cai} and as elaborated in \cite{Dey}, we consider a specific linear combination  of the four-derivative
interaction terms and express the action (\ref{actiongeneral}) in a simple form :
\begin{equation}
\textit{S} = \frac{1}{16 \pi G_4}\int \mathrm{d^4}x \sqrt{-g} \ \ \bigl(R+\frac{6}{L^{2}}
-\frac{1}{4}\textit{F}_{\mu\nu}\textit{F}^{\mu\nu}+\gamma L^2 \textit{C}_{\mu \nu\rho\lambda}
\textit{F}^{\mu\nu}\textit{F}^{\rho\lambda}\bigr) ,
\label{action}
\end{equation}
where the Weyl tensor $\textit{C}_{\mu \nu\rho\lambda}$ in four spacetime dimensions is given by, 
\begin{eqnarray}
 \textit{C}_{\mu \nu\rho\lambda}=\textit{R}_{\mu \nu\rho\lambda}+{1\over 2}(g_{\mu \lambda}R_{\rho \nu}+g_{\nu \rho} 
 R_{\mu \lambda}-g_{\mu \rho}R_{\lambda \nu}-g_{\nu \lambda}R_{\rho \mu})+{1\over 6}(g_{\mu \rho}g_{\nu \lambda}-
 g_{\mu \lambda}g_{\rho \nu})R .
\label{Weyl tensor}
\end{eqnarray}
$\gamma$ is referred to as the `Weyl coupling' in the rest of the paper, and represents the effective coupling for the higher derivative interaction terms. 
The Weyl coupling term $\textit{C}_{\mu \nu\rho\lambda}\textit{F}^{\mu\nu}
\textit{F}^{\rho\lambda}$ can also be expressed as,
\begin{eqnarray}
 \textit{C}_{\mu \nu\rho\lambda}\textit{F}^{\mu\nu}\textit{F}^{\rho\lambda}=\textit{R}_{\mu \nu \rho \lambda}
\textit{F}^{\mu\nu}\textit{F}^{\rho\lambda}-2R_{\mu\nu}\textit{F}^{\mu}\hspace{0.1mm} _{\rho}\textit{F}^{\nu\rho}
+{1\over 3}R F_{\mu\nu}F^{\mu\nu} .
\label{WeylFF}
\end{eqnarray}
The variation of the action (\ref{action}) yields the following Einstein's and Maxwell's equations :
\begin{eqnarray}
R_{\mu\nu}-{1\over2}g_{\mu\nu}R-{6\over L^2}g_{\mu\nu}-T_{\mu\nu}=0 \,,
\label{EinsteinEOM}
\end{eqnarray}
\begin{eqnarray}
 \nabla_{\mu}(F^{\mu\lambda}-4\gamma L^2 C^{\mu\nu\rho\lambda}F_{\nu\rho})=0 .
 \label{MaxwellEOM}
\end{eqnarray}
with $T_{\mu\nu}$ representing the energy-momentum tensor,
\begin{eqnarray}
 && T_{\mu\nu}={1\over 2}\bigl(g^{\alpha\beta}F_{\mu\alpha}F_{\nu\beta}-{1\over4}g_{\mu\nu}F_{\alpha\beta}F^{\alpha\beta}\bigr)+
 {\gamma L^2 \over 2}\bigl[g_{\mu\nu}C_{\delta\sigma\rho\lambda}F^{\delta\sigma}F^{\rho\lambda}-6 g_{\delta\mu}
 R_{\nu\sigma\rho\lambda}F^{\delta\sigma}F^{\rho\lambda} \nonumber \\
 &&-4\nabla_{\delta}\nabla_{\rho}(F^{\rho}\hspace{0.1mm}_{\mu}F^{\delta}\hspace{0.1mm}_{\nu})+2\nabla^{\sigma}
 \nabla_{\sigma}(F_{\mu}\hspace{0.1mm}^{\rho}F_{\nu\rho})+2g_{\mu\nu}\nabla_{\sigma}\nabla_{\delta}(F^{\delta}
 \hspace{0.1mm}_{\rho}F^{\sigma\rho})-4\nabla_{\delta}\nabla_{\mu}(F_{\nu\rho}F^{\delta\rho}) \nonumber \\
 && +4R_{\delta\sigma}F^{\delta}\hspace{0.1mm}_{\mu}F^{\sigma}\hspace{0.1mm}_{\nu}+8 R_{\mu\sigma}
 F^{\sigma\rho}F_{\nu\rho}-{2\over3}R_{\mu\nu}F^{\delta\sigma}F_{\delta\sigma}-{2\over3}g_{\mu\nu}\nabla^{\rho}
 \nabla_{\rho}(F^{\delta\sigma}F_{\delta\sigma}) \nonumber \\
 && +{2\over3}\nabla_{\nu}\nabla_{\mu}(F^{\delta\sigma}F_{\delta\sigma})-{4\over3}R g^{\delta\sigma}F_{\delta\mu}
 F_{\sigma\nu}\bigr] .
\label{EnergyMomentumTensor}
\end{eqnarray}

\subsection{\textbf{Black hole solution with spherical horizon}}
Now, we want to construct a black hole solution (i.e., solution with spherical horizon) by solving eqs.(\ref{EinsteinEOM}) and (\ref{MaxwellEOM}), 
taking into account the backreaction of the $U(1)$ gauge field on the spacetime. We take the following ansatz for the metric and the gauge field as 
considered in \cite{Dey},
\begin{eqnarray}
ds^2&=&-  f(r) e^{-2\chi(r)}  dt^2+{1\over f(r)}dr^2 +r^2 (d\theta^2+\sin ^2  \theta \ d\phi^2) \,,
\label{metric ansatz}
\end{eqnarray}
\begin{equation}
 A = (\phi(r),0,0,0) \,.
 \label{A ansatz}
\end{equation}
The consideration of the backreaction of the gauge field makes the system difficult to solve analytically. Therefore, we will solve the system 
perturbatively, up to linear order in $\gamma$ and find out the metric and the gauge field. This is a caveat in our analysis that we will keep in mind,
however we comment upon this  in section 3.2 and towards the end of this paper. 

We start with the following forms for $f(r)$, $\chi(r)$ and $\phi(r)$
\begin{eqnarray}
f(r)&=&f_0(r)\bigl(1+\mathcal{F}(r)\bigr)\,,\nonumber\\
\chi(r) &=& \chi_0(r) + \chi_1(r)\,,\\
\phi(r) &=& \phi_0(r)+ \phi_1(r) .\ \nonumber
\label{perturbationansatz}
\end{eqnarray}
where $f_0(r)$, $\chi_0(r)$ and $\phi_0(r)$ are the leading order solutions representing a Reissner-Nordstr\"{o}m black hole in four dimensional AdS space, with
\begin{eqnarray}
 f_{0}(r) &=& 1-\frac{m}{r}+\frac{q^2}{r^2}+\frac{r^2}{L^2} \,,\nonumber\\
\chi_{0}(r) &=& 0 \,, \nonumber\\ 
\phi_{0}(r) &=& 2 q \left({1\over r_h}-{1\over r}\right).
\label{zeroth order soln}
\end{eqnarray}
$m$ is an integration constant related to the ADM mass ($M$) of the black hole, to be discussed in the next section, whereas the other 
integration constant $q$ is related to the total charge of the black hole, $Q={\omega_2 \over 8\pi G_4} q$, $\omega_2$ being the volume of the unit 
$2$-sphere. Also, $r_h$ in eq.(\ref{perturbationansatz}) indicates the position of the event horizon of the black hole.

Here, $\mathcal{F}(r)$, $\chi_1(r)$ and $\phi_1(r)$ are the $\mathcal{O}(\gamma)$ corrections obtained by solving 
eqs.(\ref{EinsteinEOM}) and (\ref{MaxwellEOM}) keeping terms consistently up to linear order in $\gamma$,
\begin{eqnarray}
\mathcal{F}(r)&=& {\gamma \over f_0(r)} \bigl({r_h^3 k_1 \over L^2 r}+{2 h_2 q^2 \over r^2}+{2 q^2 k_2 \over r^2}-{16 q^2 \over 3r^2}
-{8 L^2 q^2 \over 3 r^4}+{10 L^2 m q^2 \over 3 r^5}-{16 L^2 q^4 \over 5 r^6}\bigr) \,,\nonumber\\
\chi_1(r)&=&  \gamma \bigl(k_2-\frac{2 L^2 q^2}{3 r^4}\bigr) \,,\\
\phi_1(r)&=& \gamma \bigl(k_3-{2 h_2 q\over r}k_4-{4 L^2 m q\over r^4}+{92 L^2 q^3 \over 15 r^5}\bigr) .\ \nonumber
\label{perturbations}
\end{eqnarray}
where $k_1$, $k_2$, $k_3$ and $k_4$ are dimensionless integration constants to be determined.
 
Following \cite{Myers} and as was done in \cite{Dey}, we determine those constants by imposing several constraints on the above equations. 
For example, we evaluate $k_2$ from the asymptotic behaviour of the black hole metric (\ref{metric ansatz}),
\begin{eqnarray}
ds^2 \vert_{r\rightarrow \infty}=- (f e^{-2\chi})_\infty dt^2+r^2 (d\theta^2+\sin ^2  \theta \ d\phi^2) .
\label{CFT metric}
\end{eqnarray}
where $(f e^{-2\chi})_\infty=\lim_{r\to\infty}f(r) e^{-2\chi(r)}$. This asymptotic form represents the background metric for the dual boundary CFT.
Since, the speed of light in the CFT should be unity, we demand that $(f e^{-2\chi})_\infty={r^2 \over L^2}$, which in turn gives $k_2=0$.

We evaluate $k_4$ by requiring that the charge density $q$ remains fixed. Note that, we can always write the Maxwell equation, 
eq.(\ref{MaxwellEOM}) in the form $\nabla_{\mu}X^{\mu\lambda}=0$, where, $X^{\mu\lambda}$ is an antisymmetric tensor. Hence, the 
dual of $(*X)_{\theta \phi}$ is a constant and it is appropriate to choose this constant to be the fixed charge density $q$, 
i.e., $(*X)_{\theta \phi}=q$. Since the quantity $(*X)_{\theta \phi}$ does not depend on $r$, we demand 
\begin{eqnarray}
\lim_{r\rightarrow\infty}\left(*X\right)_{\theta \phi}=q .
\label{constraint1a}
\end{eqnarray}
On the other hand, computation of this quantity in the asymptotic limit gives,
\begin{eqnarray}
\lim_{r\rightarrow\infty}\left(*X\right)_{\theta \phi}&=&\lim_{r\rightarrow\infty}\bigl[r^2 \sin \theta e^{\chi(r)}
\left(F_{r t}-8\gamma L^2 C_{r t}\hspace{0.1mm}^{r t}F_{r t}\right)\bigr] \nonumber\\
&=&\bigl(1 + \gamma k_4 \bigr)q .
\label{constraint1b}
\end{eqnarray}
A comparison of eqs.(\ref{constraint1a}) and (\ref{constraint1b}) yields $k_4=0$.

Next, we determine $k_1$ by imposing the condition, $f_0(r) \mathcal{F}(r) |_{r=r_h}=0$, which simply means that we fix the position of the event horizon at
$r=r_h$. Thus we obtain $k_1$ as, 
\begin{eqnarray}
k_1=-\frac{2 L^4 m^2}{15 r_h^6}-\frac{2 L^4 m}{5r_h^5}+\frac{8 L^4}{15 r_h^4}+\frac{34 L^2 m}{15r_h^3}-\frac{8 L^2}{5 r_h^2}-\frac{32}{15} .
 \label{constraint2}
\end{eqnarray}
The remaining constant $k_3$ is determined by demanding $A_t$ to be vanished at the horizon. This is required to have a well defined one-form for the
gauge field $A$. This condition implies $\phi_1(r_h)=0$, which in turn evaluates the constant $k_3$ as,
\begin{eqnarray}
 k_3=\frac{4 L^2 m q}{r_h^4}-\frac{92 L^2 q^3}{15 r_h^5} .
 \label{constraint3}
\end{eqnarray}

Now we have all the integration constants in hand, and so we write down the final expressions for $\mathcal{F}(r)$, $\chi_1(r)$ 
and $\phi_1(r)$ for completeness,
\begin{eqnarray}
&&\mathcal{F}(r)= {\gamma \over f_0(r)} \Bigl(\frac{34 m}{15 r}-\frac{32 r_h^3}{15 L^2 r}-\frac{2 L^2 m^2}{15 r r_h^3}-\frac{8 r_h}{5 r} 
+\frac{8 L^2}{15 r r_h}-\frac{2 L^2 m}{5 r r_h^2} -\frac{16 q^2}{3 r^2} \nonumber \\
&& \ \ \ \ \ \ \ \ \ \ \ \ \ \ \ \ \ \ -\frac{8 L^2 q^2}{3 r^4}+\frac{10 L^2 m q^2}{3 r^5}-\frac{16 L^2 q^4}{5 r^6}\Bigr) \,,\nonumber \\
&&\chi_1(r)=  - \gamma \frac{2 L^2 q^2}{3 r^{4}} \,,\nonumber\\
&&\phi_1(r)= 4 L^2 \gamma \bigl[m q \bigl({1\over r_h^4}-{1\over r^4}\bigr)+23 q^3 \bigl({1\over 15 r^5}-{1\over 15 r_h^5}\bigr)\bigr] .
\label{Final perturbations}
\end{eqnarray}
 
The Hawking temperature of the black hole is given by
\begin{eqnarray}
 &&T={\kappa \over 2\pi}=\frac{f'(r) e^{-\chi(r)}}{4\pi}\Big \vert_{r=r_h} \nonumber \\
  && \ \ \ \ \ \ \ \ \ \ \ = e^{\frac{2   L^2 q^2}{3 r_h^4}\gamma} \Bigl[\frac{3 r_h^4+L^2 r_h^2-L^2 q^2}{4 \pi  L^2 r_h^3} - 
 \frac{2 q^2 \gamma }{3 \pi  r_h^7} \left(3 r_h^4+2 L^2 r_h^2-L^2 q^2\right)\Bigr] ,
  \label{Hawking Temp}
\end{eqnarray}
where $\kappa$ is the surface gravity. According to AdS/CFT correspondence, the Hawking temperature  represents the temperature of the dual boundary field 
theory. By setting $\gamma=0$, we get back the Hawking temperature of an RN-AdS black hole, as expected.

We also obtain a relation between the mass $(m)$ and the charge parameter $(q)$ from $f(r_h)=0$,
\begin{eqnarray}
 m=r_h+\frac{r_h^3}{L^2}+\frac{q^2}{r_h}.
\label{mass}
\end{eqnarray}
Using the gauge/gravity duality, we can also write down the chemical potential, $\Phi$, of the boundary field theory,
\begin{eqnarray}
\Phi = \lim_{r\rightarrow\infty}A_t=\lim_{r\rightarrow\infty}\phi(r)={2 q\over r_h}
+{4 L^2 q\over r_h^4} \gamma \left(m -\frac{23 q^2}{15 r_h}\right) .
\label{chemical potential}
\end{eqnarray}

\subsection{\textbf{Black brane solution with planar horizon}}
For the sake of completeness, we also mention the form of the black brane solution (i.e., solution with planar horizon) with Weyl corrections in
four spacetime dimensions. In this case, we choose the following metric and gauge field ansatz,
\begin{eqnarray}
ds^2&=& -h(r) e^{-2\zeta(r)}  dt^2+{1\over h(r)}dr^2 +{r^2\over L^2} (dx^2 + dy^2) \,,
\label{planar metric ansatz}
\end{eqnarray}
\begin{equation}
 A = (\psi(r),0,0,0) \,.
 \label{planar A ansatz}
\end{equation}
where we consider the same kind of forms of the above terms as in the spherical horizon case, i.e.,
\begin{eqnarray}
h(r)&=&h_0(r)\bigl(1+\mathcal{H}(r)\bigr)\,,\nonumber\\
\zeta(r) &=& \zeta_0(r) + \zeta_1(r)\,,\\
\psi(r) &=& \psi_0(r)+ \psi_1(r) .\ \nonumber
\label{planar perturbationansatz}
\end{eqnarray}
The leading order solutions, representing a Reissner-Nordstr\"{o}m AdS black brane, are given by,
\begin{eqnarray}
h_{0}(r) &=& \frac{r^2}{L^2}-\frac{m}{r}+\frac{q^2}{r^2} \,,\nonumber\\
\zeta_{0}(r) &=& 0 \,, \nonumber\\ 
\psi_{0}(r) &=& 2 q \bigl({1\over r_h}-{1\over r}\bigr).
\label{planar zeroth order soln}
\end{eqnarray}
Again, $\mathcal{H}(r)$, $\zeta_1(r)$ and $\psi_1(r)$ are the $\mathcal{O}(\gamma)$ corrections which can be found in the same procedure as discussed
above. They have the following forms,
\begin{eqnarray}
\mathcal{H}(r)&=& {\gamma \over h_0(r)} \left(\frac{34 m}{15 r}-\frac{32 r_h^3}{15 L^2 r}-\frac{2 L^2 m^2}{15 r r_h^3}
-\frac{16 q^2}{3 r^2}+\frac{10 L^2 m q^2}{3 r^5}-\frac{16 L^2 q^4}{5 r^6}\right) \,,\nonumber\\
\zeta_1(r)&=&  - \gamma \frac{2 L^2 q^2}{3 r^{4}} \,,\nonumber\\
\psi_1(r)&=&  4 L^2 \gamma \bigl[m q \bigl({1\over r_h^4}-{1\over r^4}\bigr)+23 q^3 \bigl({1\over 15 r^5}-{1\over 15 r_h^5}\bigr)\bigr] .\ \nonumber
\label{planar Final perturbations}
\end{eqnarray}
The Hawking temperature of this black brane is given by,
\begin{eqnarray}
 &&T = e^{\frac{2   L^2 q^2}{3 r_h^4}\gamma} \Bigl[\frac{3 r_h^4-L^2 q^2}{4 \pi  L^2 r_h^3} - 
\frac{2 q^2 \gamma }{3 \pi  r_h^7} \left(3 r_h^4-L^2 q^2\right)\Bigr] .
\label{planar Hawking Temp}
\end{eqnarray}
The chemical potential of this black brane has the same expression (\ref{chemical potential}) as the black hole solution with 
spherical horizon.

\section{\textbf{Black hole thermodynamics with Weyl correction}}

In this section, we study thermodynamics of the linear order Weyl corrected black hole geometry obtained in the previous section. For this purpose,
we first construct the on-shell action by substituting 
the form of Ricci scalar from the Einstein equation into the action (\ref{action}). A straightforward calculation in four dimensions yields,
\begin{eqnarray}
S_{on-shell}=\frac{1}{16 \pi G_4} \int d^4 x \sqrt{-g} \biggl[\frac{F_{\mu\nu}F^{\mu\nu}}{4} + \frac{6}{L^2} - 2\gamma L^2 C_{\mu\nu\rho\lambda}
F^{\mu\nu}F^{\rho\lambda} \biggr] \nonumber\\
\xrightarrow{r\rightarrow\infty}\frac{\beta \omega_2}{8 \pi G_4} \biggl(r^3-\frac{q^2}{r_h}-r_{h}^3\biggr)+
\frac{\gamma q^2}{4 \pi G_4}\biggl(\frac{1}{r_h}-\frac{q^2}{15 r_{h}^5}\biggr) .
\end{eqnarray}
Here, $\omega_2$ is the volume of unit two sphere (which we set to unity for computational purposes)
and $\beta=\frac {1}{T}$ is the inverse temperature \footnote{We have set AdS radius $L=1$, and we will use this unit throughout the paper.}. 
The on-shell action is divergent in the ${r\rightarrow\infty}$ limit.
In order to make it finite, we introduce a counter term (CT) and Gibbons-Hawking (GW) term at the boundary. 
The standard forms of these quantities are 
\begin{eqnarray}
S_{GH}=-\frac{1}{8 \pi G_4} \int d^3 x \sqrt{-\sigma} \Theta , \ \ \ \  S_{CT}=\frac{1}{16 \pi G_4} \int d^3 x \sqrt{-\sigma} \biggl(\frac{4}{L} 
+L R_3 \biggr) .
\end{eqnarray}
Here $\Theta$ is the trace of the extrinsic curvature, $R_3$ is the Ricci scalar constructed from three dimensional boundary metric and $\sigma$ 
is the induced metric on the boundary.
One can check that the total action $S_{Total}= S_{on-shell} + S_{GH} + S_{CT }$ is now finite, which is given by,
\begin{eqnarray}
S_{Total} =\frac{\omega_2 \beta }{16\pi G_4 } \biggl( r_h -r_{h}^3 -\frac{q^2}{r_h}  \biggr) +\frac{\omega_2 \beta \gamma q^2}{8 \pi G_4 } 
\biggl( \frac{1}{r_h} + \frac{1}{3 r_{h}^3}-\frac{q^2}{15 r_{h}^5}\biggr) .
\end{eqnarray}
Using the gauge/gravity duality, the Helmholtz or the Gibbs free energy (depending on the ensemble) can be identified as $S_{Total}$ times 
the temperature. 

Before proceeding, we want to mention here that the divergences of the on-shell action can also be removed by subtracting a reference background from it. 
For example in the grand canonical ensemble, a pure AdS spacetime can be used as a reference background, and for the canonical ensemble, one can 
use an extremal black hole as a reference background.
This gives a definition of the on-shell action relative to that of the reference background. This reference background subtraction method 
is  perfectly okay to work with. We have explicitly checked that both regularization procedures give same form for the Gibbs free energy.

\subsection{\textbf{First law of Weyl corrected black hole thermodynamics}}

In this section, we will verify the first law of thermodynamics for linear order Weyl corrected black hole geometry. Here, we will mostly concentrate on 
the horizon with spherical topology but our results can be straightforwardly generalised to planar horizons. In order to verify the first law, 
we first need to calculate mass of the black hole. To obtain the expression for the mass, we use a prescription due to 
Ashtekar, Magnon and Das (AMD) \cite{AshtekarDas}. The AMD prescription gives a procedure to calculate a conserved quantity $Q[K]$ 
associated with a  Killing field $K$ in an asymptotically AdS spacetime as\footnote{We refer the reader to \cite{AshtekarDas} for details of the analysis.}
\begin{equation}
Q[K]=\frac{1}{8 \pi (D-3)G_4} \oint \tilde{\epsilon}^{\mu}_{ \ \nu}  K^{\nu} d\tilde{\varSigma}^{\mu} .
\label{charge}
\end{equation}
where $\tilde{\epsilon}^{\mu}_{ \ \nu}=\Omega^{D-3} \tilde{n}^\rho \tilde{n}^\sigma \tilde{C}^{\mu}_{\ \ \rho \nu \sigma}$, $\tilde{n}^\rho$ is the
unit normal vector, $\tilde{C}^{\mu}_{\ \ \rho \nu \sigma}$ is the Weyl tensor constructed from $\tilde{ds^2}=\Omega^2 ds^2 $ and $K^{\nu}$ is the
conformal killing vector field. Also $\Omega=1/r$, $D$ 
is the number of bulk spacetime dimensions and $d\tilde{\varSigma}^{\mu}$ is the area element of the $D-2$ dimensional transverse section of the AdS 
boundary. For a timelike killing vector, after some algebra, we arrive at the following expression for the conserved mass
\begin{equation}
Q[K]=M=\frac{\omega_{D-2}}{8 \pi (D-3) G_4} \Omega^{3-D} (\tilde{n}^{\Omega})^{2}\tilde{C}^{t}_{\ \ \Omega t \Omega} .
\end{equation}
specializing to $D=4$ and converting back to  $r=1/\Omega$, we get
\begin{equation}
M=-\frac{\omega_{2}}{8 \pi G_4} \frac{r^3}{3} \biggl(\frac{g_{rr} g_{tt}''}{2 g_{tt}} - \frac{g_{rr} g_{tt}'^2}{4 g_{tt}^2} + \frac{g_{rr}' g_{tt}'}{4 g_{tt}} 
-\frac{g_{rr} g_{tt}'}{2 r g_{tt}} -\frac{g_{rr}'}{2 r} -\frac{(\lambda-g_{rr})}{r^2}\biggr) .
\end{equation}
here the prime denotes the derivative with respect to $r$, $\lambda=1 \ (0)$ for spherical (planner) horizon and 
$$g_{tt}=f(r)(1+ \mathcal{F}(r))e^{-2 \chi(r)}, \ \ g_{rr}=f(r)(1+\mathcal{F} (r))$$
Now for spherical horizon, substituting the forms of $f(r)$, $\mathcal{F} (r)$ and $\chi(r)$, we get
\begin{equation}
M=\frac{\omega_{2}}{8 \pi G_4} \biggl(\frac{q^2}{r_h} +r_h +r_{h}^3\biggr)+\frac{\omega_{2} \gamma q^2}{4 \pi G_4} \biggl(\frac{q^2}{15 r_{h}^5} +\frac{1}{3 r_{h}^3}
-\frac{1}{ r_{h}}\biggr) .
\label{AMDmass}
\end{equation}
Notice from eq.(\ref{AMDmass}) that, for $\gamma=0$ case, our result for the mass reduces to that of the standard ADM mass of RN-AdS black holes in 
four dimensions, as expected. Similarly we can calculate the entropy of the black hole using the Wald formula,
\begin{eqnarray}
S_{Wald}=-2\pi \int d^2 x \sqrt h \frac{\partial \mathcal{L}}{\partial R_{\mu\nu\rho\lambda}} \varepsilon_{\mu\nu} \varepsilon_{\rho\lambda}
\end{eqnarray}
Here $\mathcal{L}$ is the Lagrangian, $\epsilon_{\mu\nu}$ is the binormal killing vector normalised by $\varepsilon_{\mu\nu}\varepsilon^{\mu\nu}=-2$ 
and $h$ is the determinant of the two-sphere metric. At the leading order in $\gamma$, for our four dimensional set up, we get
\begin{eqnarray}
&& S_{Wald}=-\frac{1}{8 G_4} \int d^{2}x \sqrt h \bigg[\biggl(1+\frac{\gamma F_{\alpha\beta}F^{\alpha\beta}}{3}\biggr)
g^{\mu\rho}g^{\nu\lambda}\varepsilon_{\mu\nu}\varepsilon_{\rho\lambda}
+ \gamma F^{\mu\nu}F^{\rho\lambda}\varepsilon_{\mu\nu}\varepsilon_{\rho\lambda} \nonumber\\
&& \ \ \ \ \ \ \ \ \ \ + 2\gamma g^{\mu\nu}F^{\sigma\rho}F_{\rho}^{\ \lambda}\varepsilon_{\sigma\mu}\varepsilon_{\lambda\nu} \biggr] \nonumber\\
&& \ \ \ \ \ \ \ \ \ \ =\frac{\omega_2 r_{h}^2}{4 G_4}-\frac{2\omega_2 \gamma q^2}{3 r_{h}^2 G_4} .
\label{WaldEntropy}
\end{eqnarray}
Finally, using the forms of potential $(\Phi)$ and charge $(Q)$ we get Gibbs Free energy as
\begin{equation}
G=M-T S -Q \Phi= \frac{\omega_2 }{16\pi G_4 } \biggl( r_h -r_{h}^3 -\frac{q^2}{r_h}  \biggr) +\frac{\omega_2 \gamma q^2}{8 \pi G_4 } 
\biggl( \frac{1}{r_h} + \frac{1}{3 r_{h}^3}-\frac{q^2}{15 r_{h}^5}\biggr) .
\label{1stlaw}
\end{equation}
which is nothing but $\frac{S_{Total}}{\beta}$. This shows that the first law of black hole thermodynamics is indeed satisfied in Weyl corrected black hole geometry,
to linear order in the Weyl coupling. 

\subsection{Thermodynamics with spherical horizon}
\begin{figure}[t!]
\begin{minipage}[b]{0.5\linewidth}
\centering
\includegraphics[width=2.8in,height=2.3in]{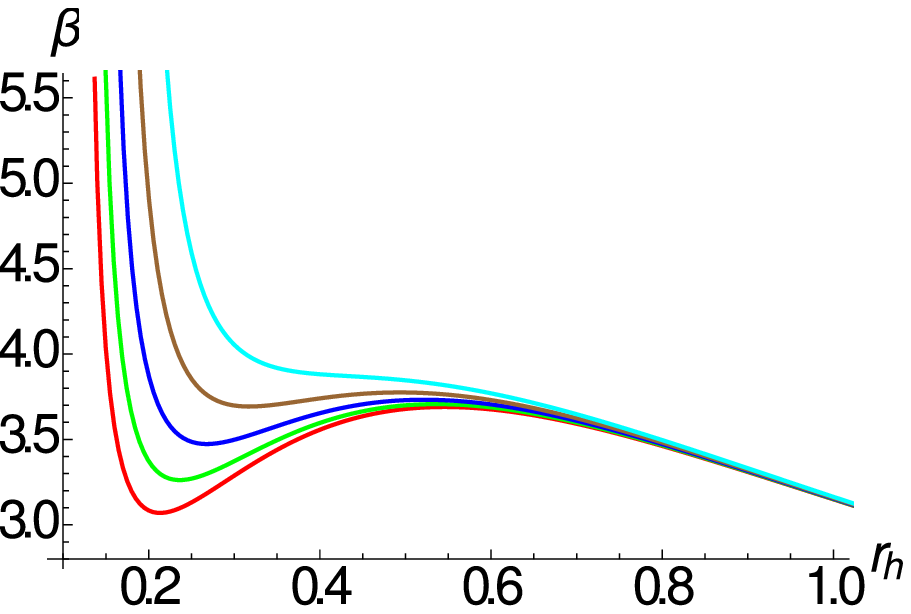}
\caption{\small $\beta$ as a function of horizon radius in four dimensions with spherical horizon. Red, green, blue, brown and cyan curves correspond to 
${1\over10}$, ${1\over 9}$, ${1\over 8}$, ${1\over 7}$ and ${1\over6}$
respectively. Here $\gamma=0.002$ is kept fixed.}
\label{4DsphericalbetavsRhgamma1by500vsq}
\end{minipage}
\hspace{0.4cm}
\begin{minipage}[b]{0.5\linewidth}
\centering
\includegraphics[width=2.8in,height=2.3in]{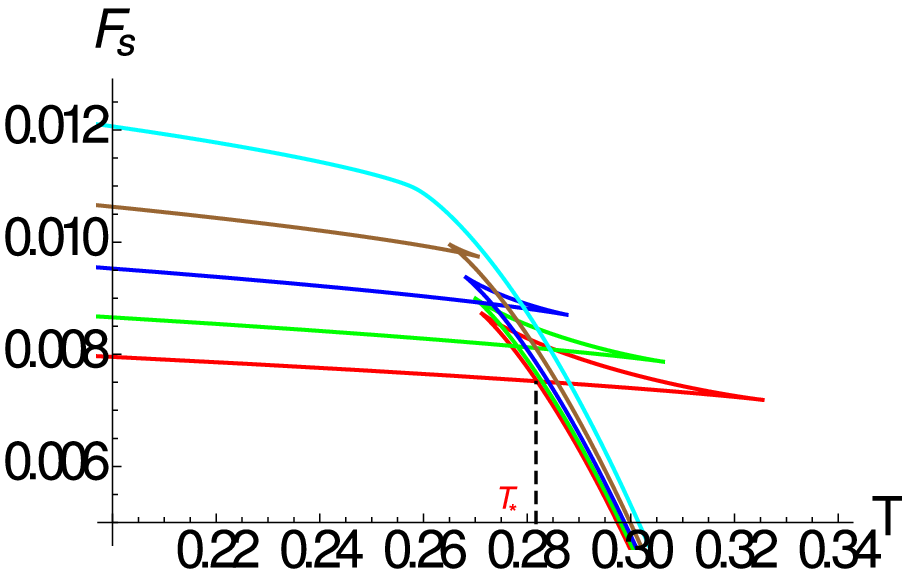}
\caption{\small Free energy as a function of temperature in four dimensions with spherical horizon. Red, green, blue, brown and cyan curves correspond to 
${1\over10}$, ${1\over 9}$, ${1\over 8}$, ${1\over 7}$ and ${1\over6}$
respectively. Here $\gamma=0.002$ is kept fixed.}
\label{4DsphericalFvsTgamma1by500vsq}
\end{minipage}
\end{figure}
\begin{figure}[t!]
\begin{minipage}[b]{0.5\linewidth}
\centering
\includegraphics[width=2.8in,height=2.3in]{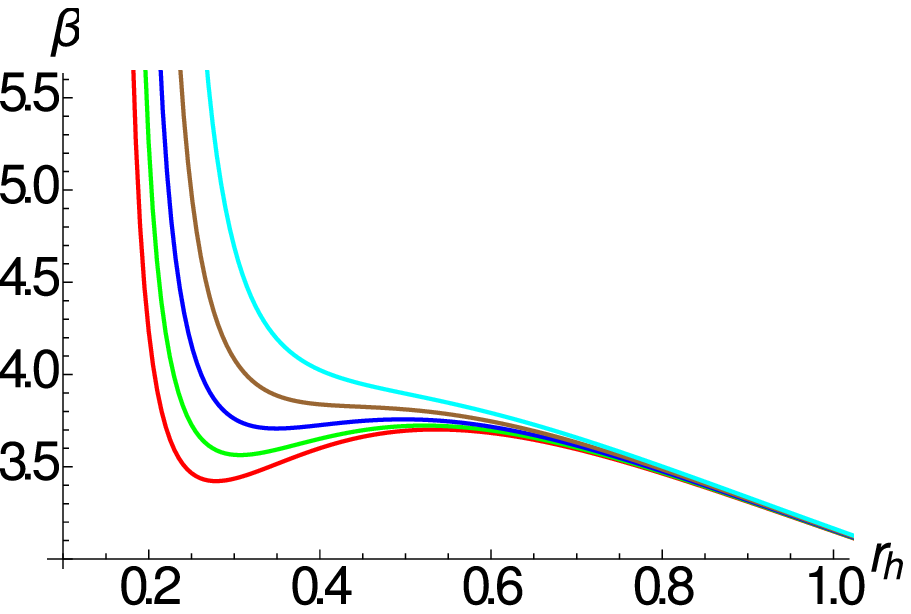}
\caption{\small $\beta$ as a function of horizon radius in four dimensions with spherical horizon. Red, green, blue, brown and cyan curves correspond to 
${1\over10}$, ${1\over 9}$, ${1\over 8}$, ${1\over 7}$ and ${1\over6}$
respectively. Here $\gamma=0.01$ is kept fixed.}
\label{4DsphericalbetavsRhgamma1by100vsq}
\end{minipage}
\hspace{0.4cm}
\begin{minipage}[b]{0.5\linewidth}
\centering
\includegraphics[width=2.8in,height=2.3in]{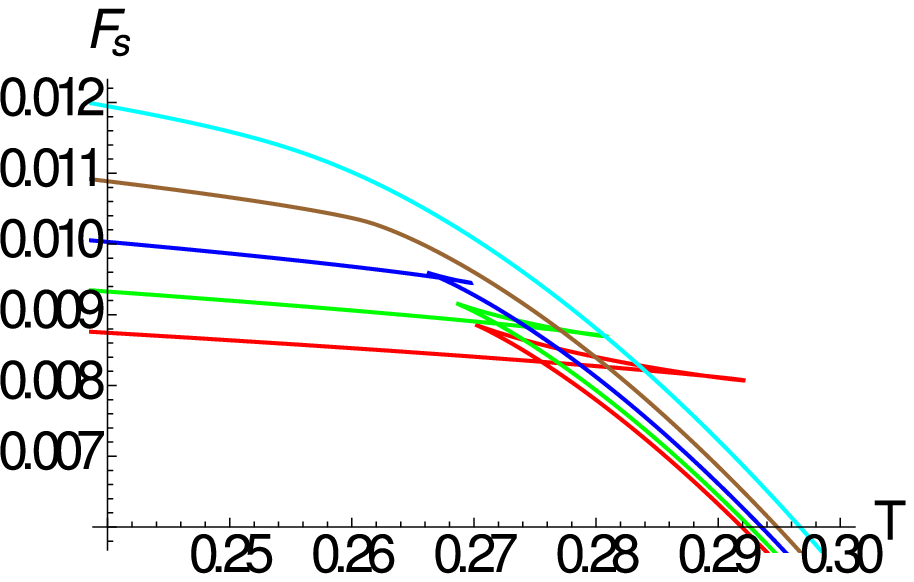}
\caption{\small Free energy as a function of temperature in four dimensions with spherical horizon. Red, green, blue, brown and cyan curves correspond to 
${1\over10}$, ${1\over 9}$, ${1\over 8}$, ${1\over 7}$ and ${1\over6}$
respectively. Here $\gamma=0.01$ is kept fixed.}
\label{4DsphericalFvsTgamma1by100vsq}
\end{minipage}
\end{figure}

After explicitly verifying the first law of thermodynamics in Weyl corrected black hole geometry, we now move on to discuss its thermodynamic properties 
in various ensembles. Here, we will mostly concentrate on the canonical ensemble (for spherical horizons), as this is the ensemble which exhibits many 
interesting features. Similar analysis for the grand canonical ensemble can be straightforwardly carried out, but we will not discuss it here.
\\
The Gibbs free energy at the linear order in Weyl coupling $\gamma$ was found in the previous subsection (eq.(\ref{1stlaw})). 
Similarly, we find the Helmholtz free energy at the leading order in $\gamma$ as
\begin{eqnarray}
F_s=G_s +Q \Phi =\frac{\omega_2 }{16\pi G_4 } \biggl(\frac{3 q^2}{r_h} + r_h -r_{h}^3 \biggr) 
+\frac{\omega_2 \gamma q^2}{8 \pi G_4 } \biggl( \frac{5}{r_h} + \frac{13}{3 r_{h}^3}-\frac{11 q^2}{5 r_{h}^5}\biggr) .
\end{eqnarray}
In the above equation, a subscript `$s$' is used to indicate that this equation is for spherical horizon.
Let us first discuss the thermodynamics of Weyl corrected geometry in the canonical ensemble by fixing the charge of the system. In order
to do this, we will need to choose numerical values for $\gamma$ (and $q$). Admittedly, in the absence of a controlled perturbative
expansion, one might question the validity of the ``smallness'' of $\gamma$.  While this is certainly a caveat, our numerical results should
be considered as illustrations of the deviations from RN-AdS behaviour up to leading order in $\gamma$, and as we show in the next section,
these are indeed consistent with results from the boundary theory. 

In fig.(\ref{4DsphericalbetavsRhgamma1by500vsq}), we have shown the variation of the inverse temperature $\beta$ with respect to 
the horizon radius $r_h$ for fixed $\gamma=0.002$. 
Here, the red, green, blue, brown and cyan curves correspond to $q$ = ${1\over 10}$, ${1 \over9}$, ${1\over 8}$, ${1\over 7}$ 
and ${1\over6}$ respectively. One can observe that for small value of $q$, say $q = {1\over 10}$, there are three branches: two stable and 
one unstable. The two stable branches, i.e., when the slope is negative, correspond to small and large black holes. These two stable branches are 
connected by an unstable black hole branch where the  slope is positive. 

The $\beta$ vs $r_h$ behaviour therefore shows that there are black holes at 
all temperatures. It also indicates a possible first order phase transition from a small black hole to a large black hole as we increase the 
temperature. The first order phase transition is confirmed by calculating the Helmholtz free energy as a function of the temperature. This is shown 
in fig.(\ref{4DsphericalFvsTgamma1by500vsq}), where swallow tail like behaviour for $q = {1\over 10}$ is apparent.
In fig.(\ref{4DsphericalFvsTgamma1by500vsq}), the unstable black hole branch makes the base of the tail. As we increase the temperature, a phase 
transition from a small black hole to a large black hole takes place at the kink, where the free energy of the large black hole becomes lower than that 
of the small black hole. The kink therefore defines a critical temperature $T_*$. For $q = {1\over 10}$, $T_*$ is shown by a dashed black line in 
figure (\ref{4DsphericalFvsTgamma1by500vsq}), we find $T_* \approx 0.2817$ in this case. 

However for larger values of $q$, say $q = {1\over 6}$, the unstable branch completely disappears in fig.(\ref{4DsphericalbetavsRhgamma1by500vsq}) 
and the two stable branches merge into a single stable black hole branch where the slope is always negative. This is also confirmed by a free energy 
calculation which shows that as we increase the charge of the black hole the swallow tail structure starts decreasing in size and completely disappears
at some critical $q_{c}$. Above $q_c$ a stable black hole branch is always present. Our analysis indicates that there exist a line of first order 
phase transition between black holes of different sizes that terminates at a second order point. The second order point is defined by $q_c$.
This is analogous to the classic Van der Waals liquid-gas systems which was first discovered in \cite{Chamblin}. Below, we will explicitly analyze
the relation between $q_c$ and $\gamma$.  

We have analyzed the phase structure of Weyl corrected black holes for other values of $\gamma$ as well. This is shown in 
figs.(\ref{4DsphericalbetavsRhgamma1by100vsq}) and (\ref{4DsphericalFvsTgamma1by100vsq}), where $\gamma=0.01$ is kept fixed. 
Here again $q$ = ${1\over 10}$, ${1 \over9}$, ${1\over 8}$, ${1\over 7}$ and ${1\over6}$ respectively, where the same colour coding as in 
figs.(\ref{4DsphericalbetavsRhgamma1by500vsq}) and (\ref{4DsphericalFvsTgamma1by500vsq}) is used. Expectedly, the essential features of our analysis are 
similar with the $\gamma=0.002$ case. However, now for ${1\over 7}$ (brown curve) we don't find any first order phase transition from a small black 
hole to a large black hole. This indicates that the higher value of $\gamma$ decreases the magnitude of critical $q_c$. We have checked this for 
several values of $\gamma$.

A word regarding the relation between $q_c$ and $\gamma$ is in order. The critical value of $q$ where small and large black hole merge together define 
an inflection point in the $\beta-r_h$ plane (or similarly in the $\beta-S_{Wald}$ plane). At this inflection point, $q_c$ and critical $r_h$ can be 
determined by the following equations,
\begin{eqnarray}
\biggl(\frac{\partial T}{\partial S_{Wald}}\biggr)_q=0, \ \ \ \biggl(\frac{\partial^2 T}{\partial S_{Wald}^2}\biggr)_q=0 .
\end{eqnarray}
At first sight, due to the complicated nature of our black hole geometry, it seems difficult to get an analytic result for $q_c$. 
However, after solving the above two equations numerically and implementing a simple fitting procedure, we obtain
\begin{equation}
q_c=0.167 +0.342 \gamma^{1/3}-1.154 \gamma^{1/2}+1.071 \gamma.
\label{qccorrected}
\end{equation}
The above equation crudely justifies our choice of ``smallness'' of the values of the Weyl coupling constant, since for the values of $\gamma$ 
that we have chosen, $q_c$ deviates from its RN-AdS value of $\frac{1}{6}$ at most at the second decimal place.

A similar result exist for the critical horizon radius. Of course in the  $\gamma\rightarrow0$ limit we get back $q_c$ for four dimensional 
RN-AdS black hole. We have also explicitly checked that at $q=q_c$, the specific heat at constant charge defined by 
$C_q=T ({\partial S_{Wald} \over \partial T})_q$ diverges, which indicates that the phase transition at the inflection point is indeed of second order.
\begin{figure}[t!]
\begin{minipage}[b]{0.5\linewidth}
\centering
\includegraphics[width=2.8in,height=2.3in]{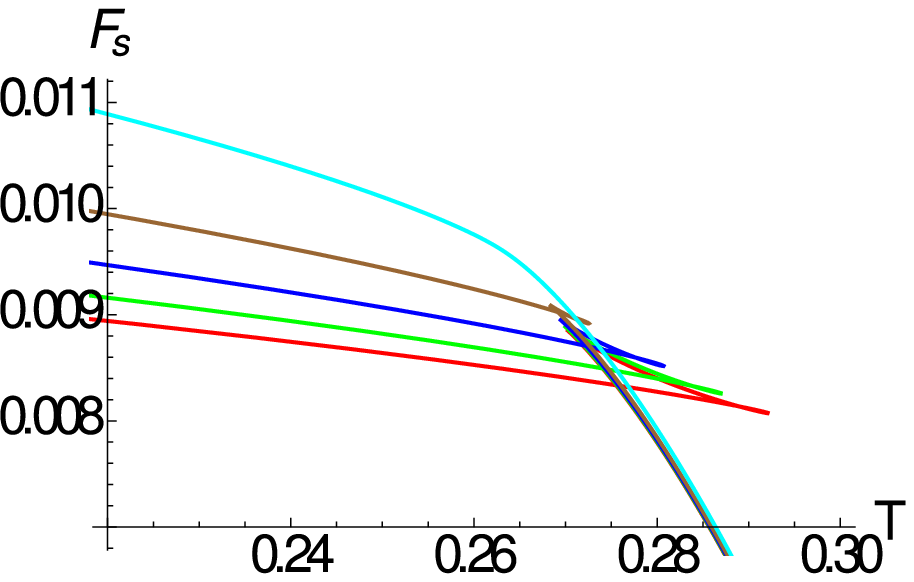}
\caption{\small $\beta$ as a function of horizon radius in four dimensions with spherical horizon. Red, green, blue, brown and cyan curves
correspond to $\gamma$ = ${1\over100}$, ${1\over 80}$, ${1\over 60}$, ${1\over 40}$ and ${1\over 20}$ respectively. Here $q={1 \over 10}$ is kept fixed.}
\label{4DsphericalbetavsRhq1by10vsgamma}
\end{minipage}
\hspace{0.4cm}
\begin{minipage}[b]{0.5\linewidth}
\centering
\includegraphics[width=2.8in,height=2.3in]{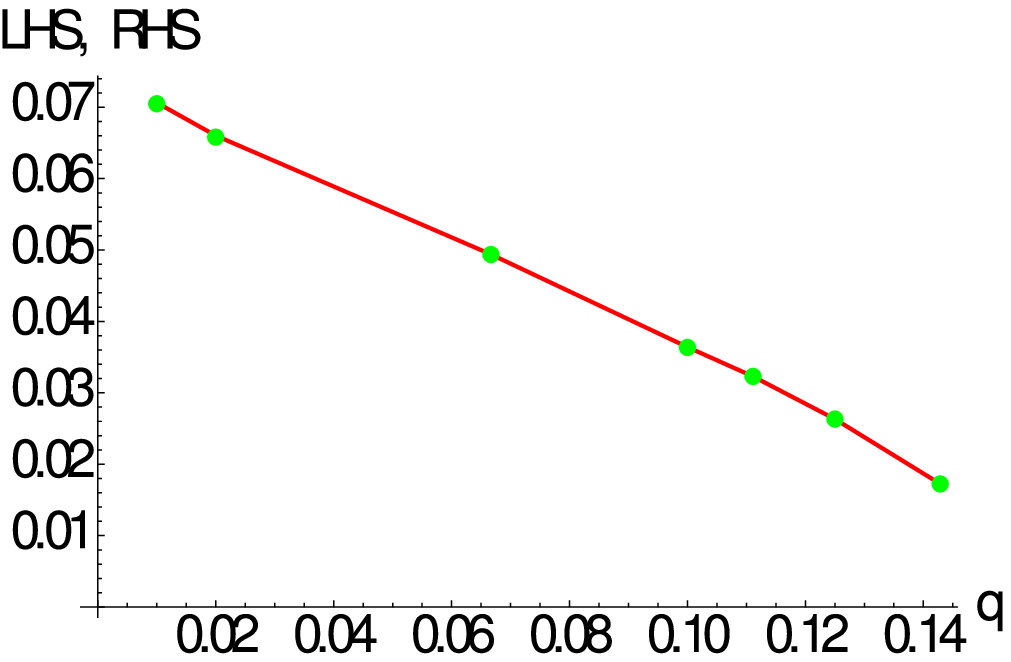}
\caption{\small Verifying Maxwell's equal area law construction. Here, we have fixed $\gamma=0.002$. Red line and green dots are the values of RHS and LHS of equation (\ref{Maxwell}) respectively.}
\label{MaxwellArea}
\end{minipage}
\end{figure}
It is important to mention that using eq.(\ref{WaldEntropy}), one can express the Hawking Temperature $T$ (eq.(\ref{Hawking Temp})) in terms
of the black hole entropy, $S_{Wald}$. It is a straightforward calculation but has a lengthy expression, so we do not write it here. Now 
performing a series expansion of this expression $T(S_{Wald})$ around the critical point $T_c$ with the critical charge $q=q_c$ one can compute the 
specific heat of the black hole from the leading order behaviour of the expansion and it turns out to be,
\begin{equation}
C_{BH}(T)=T({\partial S_{Wald}\over \partial T}) \propto (T-T_c)^{-{2\over 3}} .
\label{BHSpecificHeat}
\end{equation}
which implies that the critical exponent of specific heat in our Weyl corrected black hole geometry is same as in RN-AdS black hole case. 
In the next section, we will explicitly show that the analogous of specific heat in the context of entanglement entropy also have approximately same 
value of the critical exponent.

For completeness, in figure (\ref{4DsphericalbetavsRhq1by10vsgamma}) we have shown $F_s$ vs $T$ plot
for various values of $\gamma$ with fixed $q={1\over 10}$. Red, green, blue, brown and cyan curves correspond to 
$\gamma$ = ${1\over 100}$, ${1\over 80}$, ${1\over 60}$, ${1\over 40}$ and ${1\over 20}$ respectively. 
We again find first order like phase transition for small values of $\gamma$, which disappears at higher values.\footnote{One has to be slightly careful
here, as for higher values of the Weyl coupling, say $\gamma=0.05$, we need to keep in mind that higher order effects might become important.}  
This indicates that for a fixed charge, the Weyl coupling $\gamma$ (like the charge $Q$ for fixed $\gamma$) can equivalently control the nature of the black 
hole phase transition.  

\vspace{2mm}
Now we proceed to check Maxwell's equal area law. This amounts to verify the following equation :
\begin{equation}
T_*(S_3-S_1)=\int_{S1}^{S3} T(S,q)dS .
\label{Maxwell}
\end{equation}
where $S_3$, $S_2$ and $S_1$ correspond to the largest, intermediate and smallest roots of the equation $T(S,q)=T_*$. We have explicitly checked for
a wide range of $q$ and $\gamma$ that eq.(\ref{Maxwell}) is always satiesfied. This is shown in fig.(\ref{MaxwellArea}), where the LHS and RHS of 
eq.(\ref{Maxwell}) are plotted with respect to $q$ for $\gamma=0.002$. we see that there is indeed an excellent match.
\begin{figure}[t!]
\begin{minipage}[b]{0.5\linewidth}
\centering
\includegraphics[width=2.8in,height=2.3in]{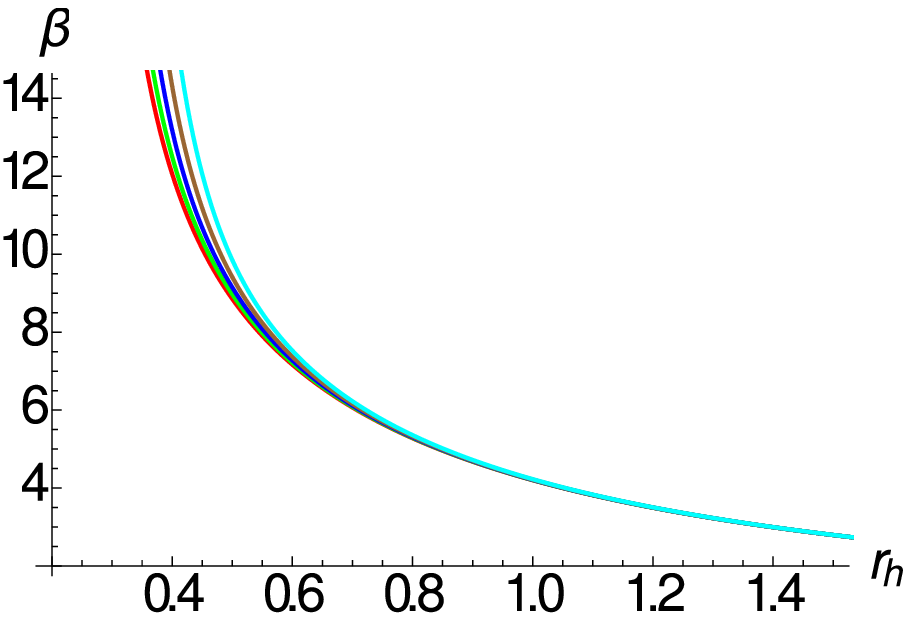}
\caption{\small $\beta$ as a function of horizon radius in four dimensions with planner horizon. Red, green, blue, brown and cyan curves correspond to 
${1\over10}$, ${1\over 9}$, ${1\over 8}$, ${1\over 7}$ and ${1\over6}$
respectively. Here $\gamma=0.002$ is kept fixed.}
\label{4DplannerFvsTgamma1by500vsq}
\end{minipage}
\hspace{0.4cm}
\begin{minipage}[b]{0.5\linewidth}
\centering
\includegraphics[width=2.8in,height=2.3in]{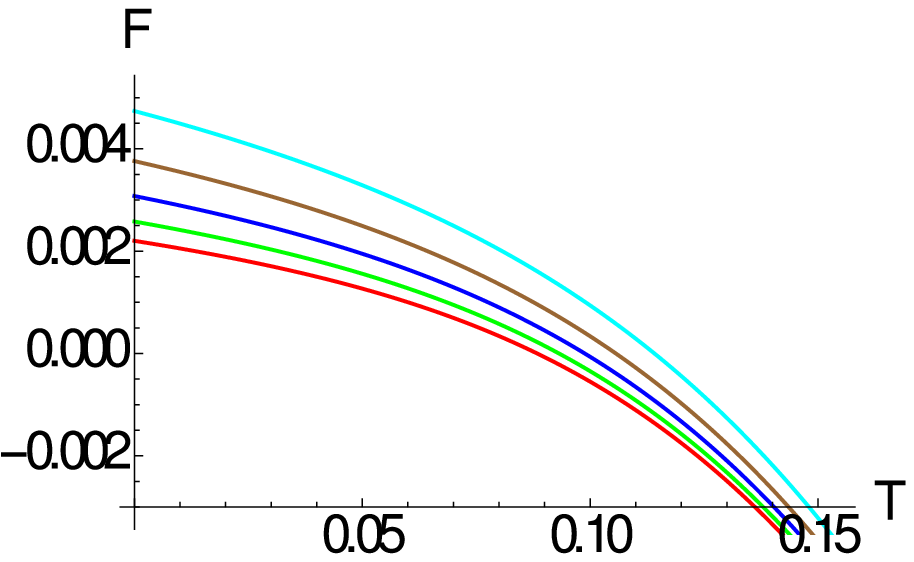}
\caption{\small Free energy as a function of temperature in four dimensions with planner horizon. Red, green, blue, brown and cyan curves correspond to 
${1\over10}$, ${1\over 9}$, ${1\over 8}$, ${1\over 7}$ and ${1\over6}$
respectively. Here $\gamma=0.002$ is kept fixed.}
\label{4DplannerbetavsRhgamma1by500vsq}
\end{minipage}
\end{figure}

\subsection{Thermodynamics with planner horizon}
We now proceed to examine the thermodynamics of Weyl corrected geometry with a planner horizon. As opposed to the spherical horizon case, 
we did not find much interesting physics in this case, and therefore we will be very brief here. Let us first record the expressions for the Gibbs and 
Helmholtz free energies at the linear order in $\gamma$,
\begin{eqnarray}
G_p= \frac{V_2}{16 \pi G_4} \biggl(-\frac{q^2}{r_h} - r_{h}^3\biggr) +\frac{V_2 \gamma q^2}{8 \pi G_4} \biggl( \frac{1}{r_h}-\frac{q^2}{15 r_{h}^5}\biggr) .
\end{eqnarray}
\begin{eqnarray}
F_p= \frac{V_2}{16 \pi G_4} \biggl(\frac{3 q^2}{r_h} - r_{h}^3\biggr) +\frac{V_2 \gamma q^2}{8 \pi G_4} \biggl( \frac{5}{r_h}-\frac{11 q^2}{5 r_{h}^5}\biggr) .
\end{eqnarray}
here $V_2$ is the volume of the two dimensional plane, and the subscript `$p$' is used to indicate that these 
expressions are strictly for planner horizon topology. In figs.(\ref{4DplannerFvsTgamma1by500vsq}) and (\ref{4DplannerbetavsRhgamma1by500vsq}),
plots for $\beta$ vs $r_h$ and $F_p$ Vs $T$ for fixed $\gamma=0.002$ in the canonical ensemble are shown. The red, green, blue, brown and cyan curves 
correspond to $q$ = ${1\over 10}$, ${1 \over9}$, ${1\over 8}$, ${1\over 7}$ and ${1\over6}$ respectively. As in the spherical horizon case, we again find 
stable black hole solutions at all the temperatures. However, there are no phase transitions here. We have 
examined the phase structure for several values of $\gamma$ and $q$, and find no qualitative differences with the ones shown in
figs.(\ref{4DplannerFvsTgamma1by500vsq}) and (\ref{4DplannerbetavsRhgamma1by500vsq}).

\section{Holographic Entanglement Entropy}
In this section, we will compute the holographic entanglement entropy for the Weyl-corrected black hole (spherical horizon) in four 
dimensions. This is relevant for the dual $3+1$ CFT. 
The purpose of this section is to show that the HEE captures all information regarding phase transition, obtained from the study of (the bulk)
black hole thermodynamics, in the Weyl corrected scenario. As pointed out in \cite{Johnson}, such an analysis of the HEE is an indicator
of finite volume phase transitions at large $N$, in the boundary gauge theory. 

Before presenting the results, let us briefly review the salient features of holographic entanglement entropy that we will need. As mentioned in the introduction, if a 
quantum system is divided into two subsystems, $\mathcal{A}$ and its complement $\mathcal{B}$, the entanglement entropy gives a quantitative measure 
of the correlation between these subsystems. The entanglement entropy of the subsystem $\mathcal{A}$ is defined as,
\begin{eqnarray}
 S_{\mathcal{A}}=-Tr_{\mathcal{A}}(\rho_{\mathcal{A}} \ln \rho_{\mathcal{A}}) .
 \label{EE}
\end{eqnarray}
where $\rho_{\mathcal{A}}$ is the reduced density matrix of $\mathcal{A}$, obtained by considering the trace over the degrees of freedom 
of $\mathcal{B}$, i.e., $\rho_{\mathcal{A}}=Tr_{\mathcal{B}}(\rho)$, where $\rho$ is the density matrix of the full quantum system. While direct
calculation of entanglement entropy in a quantum field theory is very difficult beyond $1+1$ dimensions, it becomes easy using the  
recently proposed Ryu-Takayanagi formula. Using this holographic formula, the entanglement entropy of the subsystem $\mathcal{A}$ 
living on the boundary of the AdS space is given by,
\begin{eqnarray}
S_{\mathcal{A}}={\mbox{Area} (\Gamma _{\mathcal{A}})\over 4G_N} .
\end{eqnarray}
where $G_N$ is the gravitational constant of the bulk gravity and $\Gamma _{\mathcal{A}}$ is a codimension-2  minimal area 
hypersuface which extends into the bulk and shares the same boundary $\partial{\mathcal{A}}$ of the subsystem $\mathcal{A}$.
As we have already mentioned, this formula is similar to the Bekenstein-Hawking entropy formula for black holes. 

But the Ryu-Takayanagi conjecture only holds for static backgrounds with Einstein gravity as the bulk action. 
In the presence of higher derivative terms, this conjecture may no longer hold. Here, one may think of the Wald entropy as a good guess to
prescribe an expression for the HEE for a generic gravity theory. However, this guess turns out to be wrong, since it produces erroneous universal 
terms in the expression of entanglement entropy \cite{Myers1}, \cite{Dong}. In \cite{Myers1}, it was explicitly shown that for Lovelock  gravity theory, instead
of taking the Wald entropy, if one consider the expression of Jacobson-Myers entropy \cite{JacobsonMyers}, we would get the correct universal terms in 
the entanglement entropy. The Jacobson-Myers entropy differs from Wald entropy only by terms involving the extrinsic curvature. If we have a Killing
horizon, this extra term vanishes and the two entropy formulas give the same result. 

The work of \cite{Dong} derived a general holographic
entanglement entropy formula for a general higher derivative gravity theory. It consists of Wald's entropy as the leading term and corrections due to
extrinsic curvature as subleading ones. One can think of these subleading terms as anomaly terms in the variation of the action. Since we are dealing with 
a four dimensional bulk theory dual to a three dimensional boundary field theory, there will be no anomaly term. Moreover, one can explicitly check that, 
the four derivative interaction term in our Lagrangian (see eq.(\ref{action}) $C_{\mu\nu\rho\lambda}F^{\mu\nu}F^{\rho\lambda}$, gives a vanishing contribution 
to the anomaly term of holographic entanglement entropy (in the covariant expression of \cite{Dong}). Hence we are left with the following expression 
of holographic entanglement entropy to start with,
\begin{eqnarray}
 S_{\rm EE} = -2\pi \int d^2 x \sqrt h \frac{\partial \mathcal{L}}{\partial R_{\mu\nu\rho\lambda}} \varepsilon_{\mu\nu} \varepsilon_{\rho\lambda} .
 \label{HEEWald}
\end{eqnarray}

Now we begin our study of HEE for the Weyl-corrected black hole geometry (\ref{metric ansatz}) as the background, largely following \cite{Johnson}. 
The geometry is asymptotic to global AdS, with a boundary $\mathbb{R}\times S^2 $, product of time and a 2-sphere. Hence, at any instant, the dual field 
theory lives on the 2-sphere $S^2$ which is defined by two coordinates $\theta$ and $\phi$. Now on $S^2$, we consider the region $\mathcal{A}$ to be a 
spherical cap, separated from its complement $\mathcal{B}$ by a line of constant $\theta$ $(\theta=\theta_0)$. We will consider small values of $\theta_0$ in 
all our computations of HEE, since we want to stay away from the regime where the thermal entropy dominates. 

To compute the HEE, we have to minimize the functional in eq.(\ref{HEEWald}). This 
minimized functional in the bulk would be parameterized by the function $r(\theta)$. Note that, because of the rotational symmetry of the problem, the 
functional is independent of $\phi$. Then, up to linear order in $\gamma$, eq.(\ref{HEEWald}) can be written as,
\begin{eqnarray}
S_{\rm EE} &=& \int_{\theta=0}^{\theta_0} \!\int_{\phi=0}^{2\pi} \!\!  d\theta d\phi \, \sqrt h \frac{\partial \mathcal{L}}{\partial R_{\mu\nu\rho\lambda}} 
\varepsilon_{\mu\nu} \varepsilon_{\rho\lambda}\nonumber \\
 &=& {2\pi \over 4 G} \int_{\theta=0}^{\theta_0} \!\!  d\theta \ \ r(\theta) \sin \theta \ \left(r(\theta)^2 + {1\over f(r)} \dot r (\theta)^2\right)^{1/2} 
\left(1-{8q^2\over 3r(\theta)^4}\gamma \right) \nonumber \\
&=& {2\pi \over 4 G} \int_{\theta=0}^{\theta_0} \!\!  d\theta \ \ \mathcal{L}_{Lag} 
\label{expHEEWald}
\end{eqnarray}
where, $\dot r (\theta)={d r(\theta)\over d \theta}$ and $\mathcal{L}_{Lag}=r(\theta) \sin \theta \ \left(r(\theta)^2 + {1\over f(r)} 
\dot r (\theta)^2\right)^{1/2} \left(1-{8q^2\over 3r(\theta)^4}\gamma \right)$.
 
Now one can treat the integrand in eq.(\ref{expHEEWald}) as a Lagrangian and solve the Euler-Lagrange equation,
\begin{eqnarray}
{\partial \mathcal{L}_{Lag}\over \partial{r}}-{d\over d\theta}({\partial \mathcal{L}_{Lag}\over \partial \dot r})=0 
\label{EulerLagrange}
\end{eqnarray}
to obtain the function $r(\theta)$, with the boundary conditions ${\dot r}(0)=0$, and $r(\theta_0)\rightarrow \infty$. Since the integral is UV-divergent 
we regulate it by integrating up to a cut-off $r_c={l^2\over \xi}$ where $\xi$ is very small. We have not shown here the explicit form of the Euler-Lagrange
equation since it is lengthy, and to find out an analytical solution for $r(\theta)$ is difficult. It is a second order non-linear differential 
equation of $r(\theta)$ and we solve it numerically, using MATHEMATICA. Then we substitute the solution $r(\theta)$ and 
$\dot r(\theta)$ in eq.(\ref{expHEEWald}) to calculate the HEE.
\begin{figure}[t!]
\centering
\includegraphics[scale=1.0]{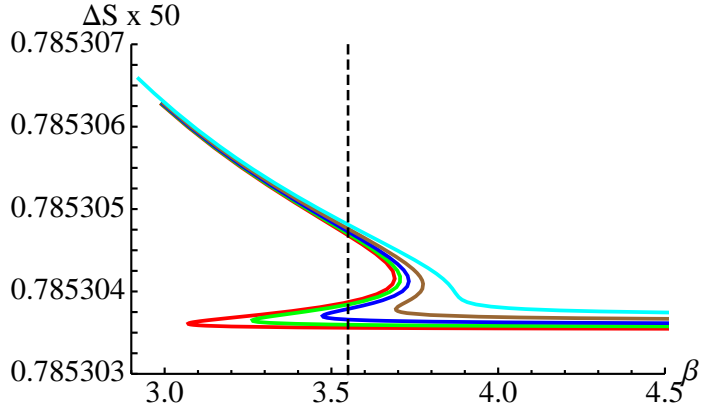}
\caption{ $\Delta S$ (scaled by $50$) as a function of $\beta$ for $\gamma=0.002$. Red, green, blue, brown and cyan curves
correspond to $q$ = ${1\over10}$, ${1\over 9}$, ${1\over 8}$, ${1\over 7}$ and ${1\over6}$ respectively. See text for more details.}
\label{DeltaSvsBetaL1GammaPt002Differentq}
\end{figure}
Note that an exact analytic solution for the Euler-Lagrange equation is possible with the pure AdS space as the background, 
i.e., with $m=0$ and $q=0$ \cite{Johnson} 
\begin{eqnarray}
r(\theta)= \bigl[\bigl({\cos(\theta)\over \cos(\theta_0)}\bigr)^2-1\bigr]^{-1/2} .
\label{solnPureAdS}
\end{eqnarray}
where, $\cos(\theta_0)={r_0 \over \sqrt{1+r_0^2}}$.
Substituting eq.(\ref{solnPureAdS}) into eq.(\ref{expHEEWald}), we have the following expression for HEE with the pure AdS background,
\begin{eqnarray}
 S_0={2\pi \over 4 G}\bigl[{1\over \xi}\bigl(1+{\xi^2}\bigr)^{1/2}\sin(\theta_0)-1\bigr] .
\end{eqnarray}
Henceforth, in all our numerical computations, we will set $G=1$, $L=1$, $\xi=10^{-4}$ and $\theta_0=0.005$. We now proceed to compute 
the HEE with the black hole background, i.e., with $m\neq 0$ and $q\neq 0$. We restrict ourselves to the case of the canonical ensemble and hence we will
fix the charge, and compute the entanglement entropy by varying the horizon radius $r_h$. The variation with $r_h$ changes
the temperature (as de[icted in figs.\ref{4DsphericalbetavsRhgamma1by500vsq} and \ref{4DsphericalbetavsRhgamma1by100vsq}) and thus we can track 
the entanglement entropy at any temperature. 

Fig.(\ref{DeltaSvsBetaL1GammaPt002Differentq}) shows the behaviour of entanglement entropy as a 
function of the inverse temperature $\beta$. Here have shown $\Delta S = S-S_0$, i.e., the entropy 
by subtracting the contribution from the Pure AdS background.
The red, green, blue, brown and cyan curves correspond to $q=$ $\frac{1}{10}$, $\frac{1}{9}$, $\frac{1}{8}$, 
$\frac{1}{7}$ and $\frac{1}{6}$ respectively. Take for example a subcritical charge, e.g., $q=\frac{1}{10}$ (the red curve). There are three branches of $\Delta S$. 
This feature ceases to be valid beyond $q_c$ computed in the last section (see eq.(\ref{qccorrected})). 
For example, at $q = 1/6$, one sees a single branch, as expected. 

For $q=\frac{1}{10}$, the HEE should have a discontinuity at the transition temperature $T_*=0.2817$ (and hence $\beta_*=3.54988$) as
follows from a bulk analysis. This is shown by the vertical black dashed line.\footnote{An analog of the Maxwell equal area law for the entanglement 
entropy \cite{Nguyen} should be valid here.} Qualitatively similar features are seen for other values of $\gamma$, and we shall not enter into the 
details here. We only note that the entanglement entropy also encodes the fact that the value of the critical charge decreases with increase of
the Weyl coupling, as expected from a bulk analysis. 

\begin{figure}[t!]
\begin{minipage}[b]{0.5\linewidth}
\centering
\includegraphics[width=2.8in,height=2.3in]{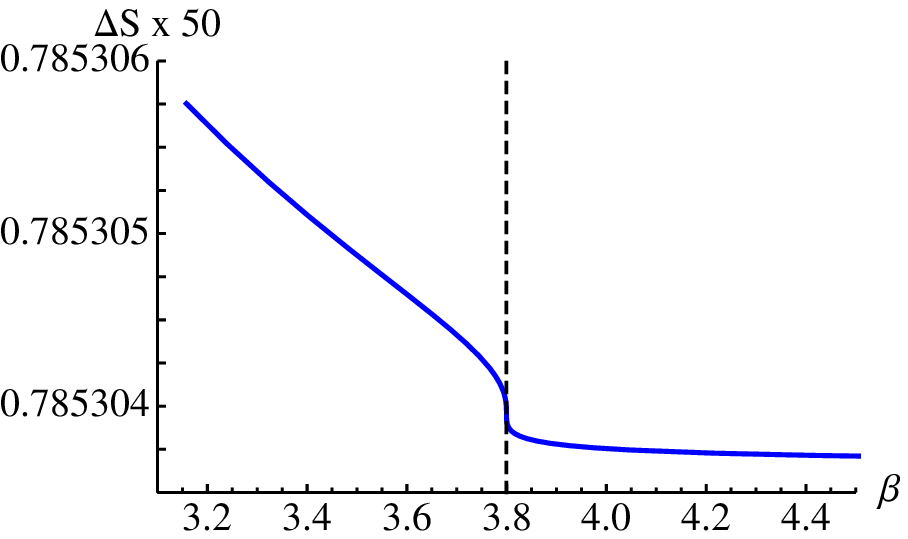}
\caption{\small $\Delta S$ (scaled by a factor of $50$) as a function of inverse temperature $\beta$ with the critical charge 
$q=q_c\approx0.137$ keeping $\gamma=0.01$ fixed. The vertical dashed line denotes the second order transition temperature, $T_c=3.79964$.}
\label{CriticalChargeCaseL1GammaPt01}
\end{minipage}
\hspace{0.4cm}
\begin{minipage}[b]{0.5\linewidth}
\centering
\includegraphics[width=2.8in,height=2.3in]{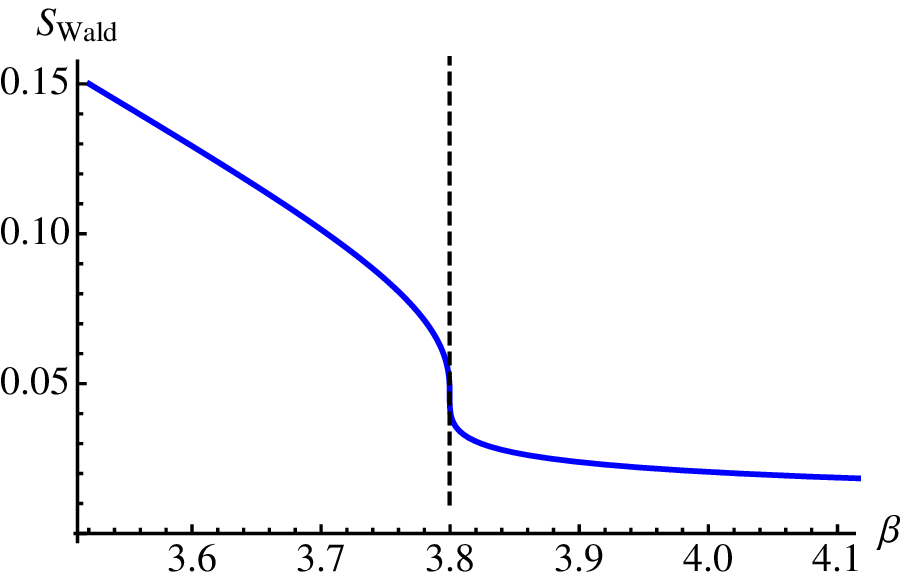}
\caption{\small $S_{Wald}$ as a function of inverse temperature $\beta$ with the critical charge $q=q_c\approx0.137$ keeping $\gamma=0.01$ fixed. 
The vertical dashed line denotes the second order transition temperature, $T_c=3.79964$. Note the similarity with fig.(\ref{CriticalChargeCaseL1GammaPt01}).}
\label{BlackHoleEntropyvsBetaCriticalChargeCaseL1GammaPt01}
\end{minipage}
\end{figure}

Now consider the second order phase transition, arising due to the critical charge $q=q_c\approx0.137$ and $\gamma=0.01$. The behaviour of the 
entanglement entropy at this transition is shown in fig.(\ref{CriticalChargeCaseL1GammaPt01}), where the two branches of the entanglement
entropy merge at $\beta_c=3.79964$. The same values for the critical charge and the critical temperature for the second order critical point is obtained 
from the bulk thermodynamics as well,
and this is depicted in fig.(\ref{BlackHoleEntropyvsBetaCriticalChargeCaseL1GammaPt01}). Indeed, an analogue of the specific heat can be 
defined for entanglement entropy, like the black hole specific heat $C_{BH}(T)$,
\begin{equation}
C_{EE}(T)=T {\partial S \over \partial T} 
\label{specificheat}
\end{equation}
We computed the entanglement entropy $S(T)$ with the critical charge $q=q_c$ in the region very close to the critical temperature $T_c$ and find 
that the resulting curve for the specific heat agrees well with the following fit,
\begin{eqnarray}
C_{EE}(T)= k \ (T-T_c)^{-0.699481} .
 \label{fitspecificheat}
\end{eqnarray}
where $k$ is a constant. This determines the critical exponent to be $\alpha=0.699481$ which is 
close to the critical exponent for the black hole specific heat, $\alpha_{BH}={2\over 3}$.

\section{Conclusions and Discussions}

In this paper, we have considered thermodynamics and entanglement entropy for four dimensional charged black holes in the presence of an extra
control parameter - the Weyl coupling constant $\gamma$, which was treated perturbatively, with our results valid up to first oder in $\gamma$. 
The following is a summary of the results obtained in this paper. 

We first computed the black hole and the black brane solutions for the Weyl corrected Einstein-Maxwell action. This was then used to compute the relevant 
thermodynamic quantities and hence the first law of thermodynamics was verified. Then, we studied black hole thermodynamics in the canonical (fixed charge)
ensemble and found the familiar Van der Waals type liquid-gas phase transition between a small and large black hole phase. We showed that for a fixed
electric charge, the control parameter $\gamma$ can be tuned to turn on a swallow tail like behaviour in the Helmholtz free energy indicating a first order
phase transition that culminates in a second order critical point. The critical exponent for the specific heat was calculated and the Maxwell's equal area construction
was also explicitly verified in this cases. 

Next, we computed the holographic entanglement entropy in the presence of a Weyl correction, and showed that it exhibits analogous behaviour as 
the Wald entropy. The specific heat was calculated from the entanglement entropy, and it showed a critical exponent very close to that
obtained from the black hole side.  
We have thus established that the Weyl coupling constant, while non-trivially modifying thermodynamic quantities, gives rise to a consistent picture in the
framework of the AdS/CFT correspondence. This complements the earlier work of \cite{Dey} in a broader framework. 

An important caveat in our analysis is that we have treated the Weyl coupling perturbatively, and hence fixed small numerical values of $\gamma$. 
It was difficult to obtain a controlled perturbation theory, which is certainly a limitation, but nonetheless we have shown that to the degree of approximation
that we have considered, the bulk thermodynamics matches exactly with boundary results. We expect the same to hold good for higher orders
in $\gamma$. 

We end by pointing out that it will be interesting
to consider modifications of Weyl-corrected black hole thermodynamics in the context of an extended phase, where the AdS radius is related to the number of 
colours $N$ of the dual gauge theory via AdS/CFT and a chemical potential conjugate to $N$ can be computed. It was recently shown in \cite{tapo} that such 
an analysis might indicate important quantum effects in a grand canonical ensemble and it will be interesting to explore this further in the context of 
Weyl-corrected gravity models.

\end{document}